# Nonlinear slow-timescale mechanisms in synaptic plasticity

Cian O'Donnell[1,2]

[1]School of Computing, Engineering, and Intelligent Systems, Magee Campus, Ulster University, Derry/Londonderry, UK.

[2]School of Computer Science, Electrical and Electronic Engineering, and Engineering Maths, University of Bristol, Bristol, UK.

**Abstract**

Learning and memory relies on synapses changing their strengths in response to neural activity. However there is a substantial gap between the timescales of neural electrical dynamics (1—100 ms) and organism behaviour during learning (seconds—minutes). What mechanisms bridge this timescale gap? What are the implications for theories of brain learning? Here I first cover experimental evidence for slow-timescale factors in plasticity induction. Then I review possible underlying cellular and synaptic mechanisms, and insights from recent computational models that incorporate such slow-timescale variables. I conclude that future progress on understanding brain learning across timescales will require both experimental and computational modelling studies that map out the nonlinearities implemented by both fast and slow plasticity mechanisms at synapses, and crucially, their joint interactions.

**Introduction**

During learning, how does any given synapse decide whether to strengthen or weaken, and by how much? One research strategy has been to try to distil the complicated mapping between neural signals and eventual synaptic strength changes into simplified 'rules of plasticity' (Figure 1). The most famous such rule comes from Donald Hebb's proposal that co-active neurons strengthen their connections, later paraphrased by Carla Shatz as 'cells that fire together wire together' [1,2]. However neurons show correlated activity on a range of timescales, from milliseconds to seconds [3]. This is faster than the typical dynamics of organism behaviour during learning, which occur over seconds—minutes [4]. Therefore a crucial unanswered question is: over what timescale of neural activity should the rules of plasticity be sensitive to? In this review I argue that there is substantial existing empirical evidence for nonlinear slow-timescale factors in plasticity, and that they should be incorporated into future theoretical models of brain learning.

The process of synaptic plasticity is commonly separated into three phases: induction, expression, and maintenance/consolidation (Figure 1). Induction is the immediate processes linking neural activity patterns to a biochemical 'decision' as to whether the synapse should strengthen, weaken, or stay the same. Expression is the physical manifestation of synaptic strength change, which may happen with a delay of several seconds to minutes following induction [5–7]. Maintenance and consolidation are the subsequent processes of storing or modifying the synaptic strength change over the long-term of hours—days, and perhaps up to years. Here I focus only on the induction phase.

Note that there are also slow heterosynaptic plasticity processes which take place over minutes—hours [8,9], and even slower ~24-hour timescale homeostatic processes like synaptic scaling [10], believed to be important for stabilising neural activity levels [11]. In this review I focus only on 'homosynaptic' plasticity induction; but how all these multi-time and spatial scale processes interact during learning is an important area of study [12].

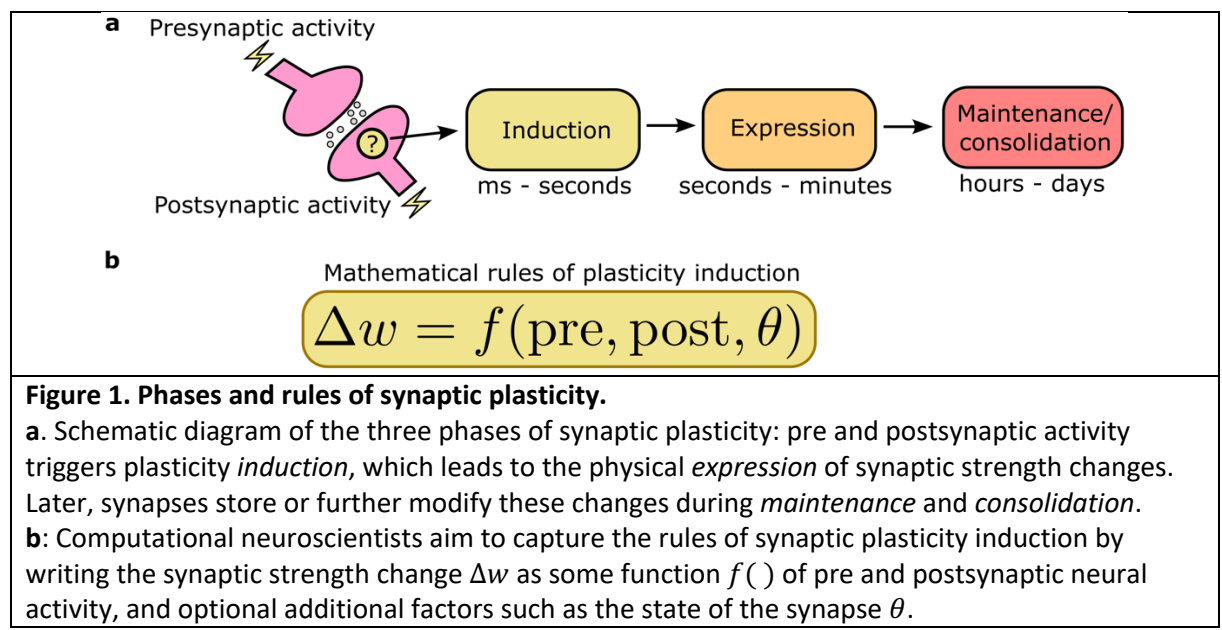

**Figure 1. Phases and rules of synaptic plasticity.**
**a**. Schematic diagram of the three phases of synaptic plasticity: pre and postsynaptic activity triggers plasticity *induction*, which leads to the physical *expression* of synaptic strength changes. Later, synapses store or further modify these changes during *maintenance* and *consolidation*.
**b**: Computational neuroscientists aim to capture the rules of synaptic plasticity induction by writing the synaptic strength change $\Delta w$ as some function $f(\ )$ of pre and postsynaptic neural activity, and optional additional factors such as the state of the synapse $\theta$.

**Experimental evidence for slow-timescale factors in synaptic plasticity induction**

Classic synaptic physiology work has long demonstrated that the duration of stimulation affects plasticity. For example in hippocampal slices, intense stimuli presented for 1—2 seconds can induce long-term potentiation (LTP), but shorter duration stimuli do not [13,14], while 5 stimulation pulses delivered over 200 ms can induce LTP whereas the same number of pulses over 40 ms do not [15]. Similarly, reliable induction of long-term depression (LTD) in *ex vivo* hippocampal slices typically requires more prolonged stimulation of at least 30 s, or even several minutes [16,17]. Together these findings demonstrated that the plasticity signalling pathways at synapses are sensitive to timescales of seconds—minutes, which is slower than the typical dynamics of neuronal electrical activity. A second implication of these experimental results is that the slow timescale factors are nonlinear: it is not the case that a 500 ms-duration stimulation induces 50% of the plasticity that a 1000 ms-duration stimulation would. Instead, some threshold duration and/or intensity of stimulation must be exceeded. We will elaborate on the importance of these nonlinearities below.

More evidence comes from spike-timing-dependent plasticity (STDP) experiments: since a single presentation of a pre-to-postsynaptic spike pairing is rarely sufficient to trigger long-term plasticity [17], experimenters typically repeat the stimulation patterns tens or hundreds of times at some regular rate, for example 1—10 times per second. This implies that STDP experiments involve three distinct timescales: most obviously the pre-post spike pairing interval (typically <50 ms), but also the repetition interval (typically 50 ms – 10 seconds), and the total stimulation duration (typically 1—10 minutes). These latter two slower seconds—minutes timescale factors, although receiving less attention, have as much effect on the direction of synaptic plasticity as the millisecond-timescale factors [18–22]. In cortical slices, fast repetition-rate stimuli induce LTP but not LTD [18] – however both LTP and LTD are seen at fast and slow repetition rates in hippocampal CA3-CA1 and corticostriatal synapses [19,23–25]. And, as with classic frequency-dependent plasticity experiments, LTP requires several repeated spike pairings, and LTD in hippocampal slices can only be induced with long-duration stimulus trains [19,22].

Further evidence for slow-timescale factors in plasticity comes from a recently discovered phenomenon termed 'behavioural time scale plasticity' (BTSP), observed in rodent hippocampus and cortex [26,27*,28,29**,30]. This non-Hebbian type of plasticity was proposed to explain the experience-dependent learning of place fields in the mouse hippocampus *in vivo*, and involves pre-synaptic input activity separated in time with post-synaptic activity with an interval of up to ~4 seconds. This seconds-timescale pre-post activity gap is slower than the sub-second dynamics of neural activity in these neurons.

Finally, there is also substantial evidence for much slower hours—days timescale factors affecting plasticity. For example, Kramar et al [31] found that if LTP was induced in rat hippocampal slices, further synaptic plasticity could not be induced until at least 60 minutes later. Similarly, experiments on the phenomenon of synaptic-tagging-and-capture [32–34] and its behavioural analogue [35] describe interactions between the plasticity induction processes at neighbouring synapses on a timescale of 1—2 hours. At even longer timescales of days to weeks, Wiegert et al [36] found that induction of LTP at synapses in *in vitro* hippocampal slice cultures reduced their ability to induce LTD 24 hours later. Cai et al [37] found that mice trained on two events separated by 5 hours had more overlapping neuronal representations in hippocampal CA1 and more linked memory recall than when the training events were spaced 7 days apart. Collectively, these examples demonstrate the existence of slow processes that mediate plasticity and learning over timescales of hours—days.

**Biological mechanisms underlying slow timescales factors**

It is useful to separate slow-timescale plasticity mechanisms into two classes: 'direct' mechanisms that inherently affect neural activity, and 'hidden' mechanisms that do not.

Direct mechanisms include short-term plasticity and post-synaptic ion channel dynamics. For example, presynaptic neurotransmitter vesicle pools tend to run down with repeated use, before recovering on a timescale of seconds [22,38]. On the postsynaptic side, a major source of depolarisation at synapses during plasticity is the back-propagating action potential (BAP) [39]. However the voltage-gated sodium channels in dendrites that support back-propagation can inactivate from prolonged depolarisation, meaning that if spikes are repeatedly evoked in the soma, the amplitude of the BAP at synapses will decrease over time [40,41]. Similarly, dendritic voltage-gated calcium channels, which affect both post-synaptic voltage dynamics and dendritic spine calcium influx, can inactivate on a timescale ~100 ms – seconds [42,43], while auxiliary AMPA receptor subunits can enable post-synaptic current dynamics in the ~100 ms range [44]. Together these seconds-timescale mechanisms affect presynaptic transmitter release and postsynaptic voltage dynamics, which in turn affect subsequent induction of long-term synaptic plasticity.

Hidden mechanisms include the many biochemical signals underlying synaptic plasticity induction: postsynaptic calcium and other second messenger signalling, downstream kinase and phosphatase activity, astrocyte interactions [45,46], protein synthesis and trafficking, and gene expression. Neural activity drives fast and large increases in postsynaptic calcium concentration, with millisecond rise-times and ~10 ms decay times [47]. Blocking these calcium increases blocks most forms of synaptic plasticity [48]. Calmodulin is one of the dominant calcium-binding proteins in neurons, binding calcium within 100 μs then releasing over ~20 ms [49]. It is also a key nonlinearity in the synaptic plasticity induction pathway: since each calmodulin molecule has two pairs of calcium-ion binding sites, the concentration of fully-bound calmodulin goes approximately as the fourth power of postsynaptic calcium concentration [49].

This calcium-bound-calmodulin (Ca-CaM) drives activity of downstream kinases and phosphatases. CaMKII is a nonlinear, leaky integrator of Ca-CaM important for LTP [50]. Its autophosphorylation mechanism drives a positive-feedback loop within each 12-subunit CaMKII macromolecule, which gives it nonlinear bistable behaviour on fast timescales [51–53**]. However its activity then decays back to baseline on a timescale of ~10 seconds [52]. This seconds-timescale sustained biochemical activity may be the key mechanism bridging the time gap between pre and postsynaptic activity in BTSP [54,55]. Other synaptic kinases may have even longer 10 minute-timescale activity [56]. Calcineurin is a phosphatase activated by Ca-CaM and is important in LTD [57]. It is not thought to be as nonlinear as CaMKII, but does have similar seconds-timescale decay kinetics [58]. Fujii et al [58] performed a clever series of experiments mapping out the relative sensitivities of both CaMKII and calcineurin to glutamate pulse stimulation frequency and number. They found that CaMKII had a supralinear dependence, maximally activated by a large number pulses delivered at high-frequencies. In contrast, calcineurin was an approximately linear integrator of pulse number, and was mostly insensitive to pulse frequency – which is a sublinearity. These differential sensitivities to input dynamics may contribute to nonlinearities in the rules of plasticity [58,59].

**Computational models of synaptic plasticity with slow variables**

Although most computational models of synaptic plasticity have a 'learning rate' parameter that is typically set to be slow relative to the timescale of neural activity, the implicit assumption is that the actual nonlinear 'decision' made by the synapse to strength or weaken is made based on fast-timescale components of neural activity, and the resulting net synaptic strength change is then linearly accumulated over longer timescales due to the slow learning rate. This formulation cannot account for the experimental data described above showing that slow-timescale factors contribute to plasticity induction.

At the other end of the scale, models have been proposed that can successfully account for slow-timescale processes in synaptic plasticity, at varying degrees of abstraction [60–66]. However these studies also tend to highly simplify and linearise the fast-timescale aspects of neural dynamics.

A key direction for future studies will therefore be to build and study computational models that include both fast and slow-timescale nonlinear components in the plasticity induction rule. Of the few multi-timescale plasticity models that do exist, most have substantially more sophisticated and data-constrained fast components than slow components [67–69]. This disparity may be due to an historic lack of quantitative data on the slow-timescale components of plasticity and their underlying mechanisms.

One exception to the above is a recent study by Rodrigues et al [70**], who built a biophysically plausible multi-timescale model of plasticity induction at the rodent hippocampal CA3-CA1 synaptic connection. The model included nonlinear dynamics of fast millisecond timescale processes such as AMPA receptor binding to glutamate and voltage-gated calcium channel switching, up to slower multi-second timescale processes such as presynaptic vesicle depletion and postsynaptic enzyme activity (Figure 2). This model could replicate plasticity experiment data from several experimental studies, including their strong nonlinear dependencies on both fast- and slow-timescale components of the induction stimuli. Interestingly, Rodrigues et al [70**] also found that plasticity in the model was sensitive to jittering of fast ~20 ms timescale components of spike train stimuli, in line with cortical data [18]. In contrast, a previous computational modelling study by Graupner et al [71] using natural spike trains recorded *in vivo* found the opposite, that even greater spike time jitter of ~80 ms had no effect on plasticity. Why the inconsistency? One possible explanation is as follows: in order for fast-timescale information in neural activity to propagate through to the slow plasticity variables without getting averaged away, the fast signal needs to be repeated consistently many times [72]. Although this condition is met in standard *in vitro* and *ex vivo* experiments, it may be less likely in aperiodic natural neural activity *in vivo* [71]. So although synapses appear to have the machinery to learn from fine-timescale information, it remains unclear whether or not they use this capability *in vivo*. The answer depends on three factors: 1) the degree of fast-timescale spike precision *in vivo*; 2) the degree of repetition of these fast signals relative to the slow variables of plasticity; 3) the alignment of both fast and slow nonlinearities in the plasticity induction cascade with the temporal statistics of information in neural activity.

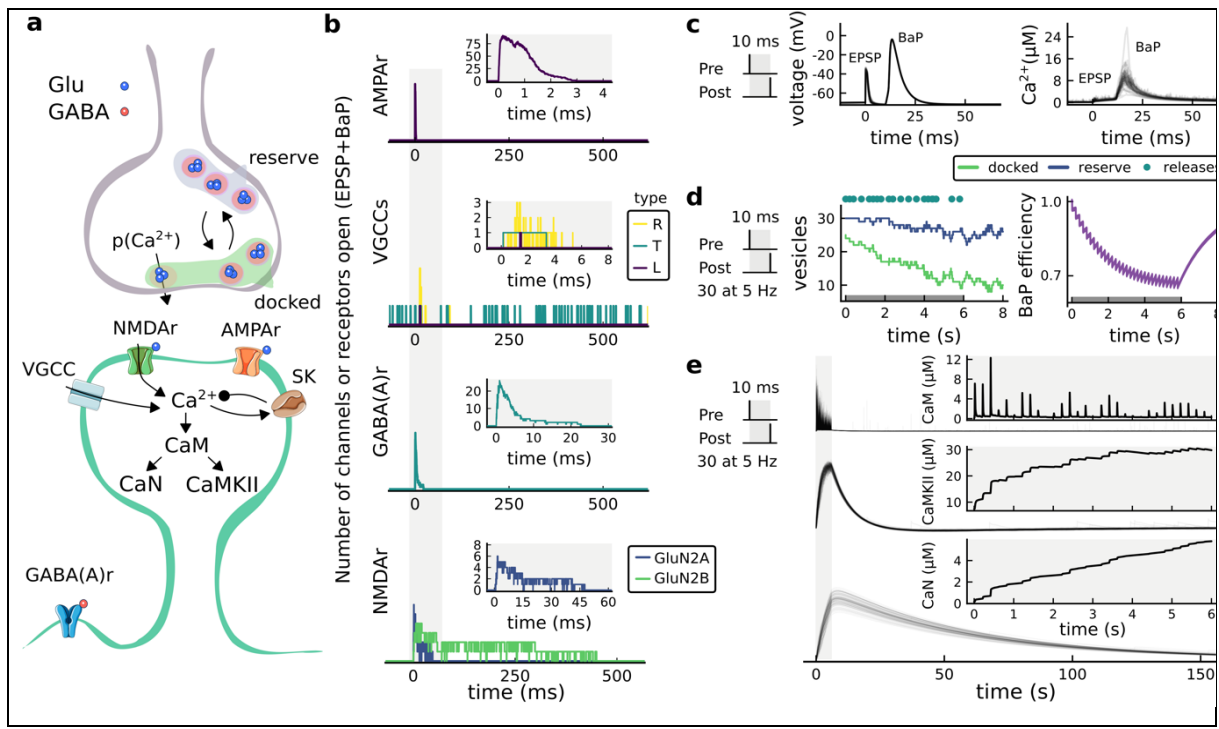

**Figure 2. Multi-timescale computational model of plasticity induction at a single hippocampal synapse by Rodrigues et al [70**].**

**a**) Schematic diagram of interactions of pre- and post-synaptic variables in the plasticity model.

**b**) Fast timescale variables AMPA and NMDA receptor binding with glutamate, voltage-gated calcium channels (VGCCs), and GABA(A) receptors in response to a single presynaptic stimulation.

**c**) Fast timescale dendritic spine voltage (left) and calcium concentration) in response to a paired presynaptic stimulation (EPSP) and postsynaptic spike (BAP).

**d**) Slow seconds-timescale variables that directly affect neural activity: presynaptic vesicle pool dynamics (left) and amplitude of back-propagating action potential due to dendritic sodium channel inactivation (right), in response to a train of 30 pre-post spike pairings.

**e**) Enzyme activity that is 'hidden' from neural activity: fast calmodulin (CaM) binding with calcium (top), ~10 second dynamics CaMKII (middle) and calcineurin (bottom), in response to a train of 30 pre-post spike pairings. Figure adapted from [70**].

**Implications and outlook**

It will be important for us not only to understand the relationship between timescales of information in neural activity and the timescales in plasticity rules, but also to understand the relationship between nonlinearities of fast and slow timescale components within the plasticity rule itself. To highlight this point, I simulated two versions of an abstract model of a plasticity induction pathway where a single fast variable (for example postsynaptic voltage or calcium) activates a single slow variable (for example CaMKII or calcineurin) (Figure 3). The slow variable in the first model was a linear function of the fast variable, whereas in the second model the slow variable was nonlinear, activated in proportion to the fourth power of the fast variable (Figure 3b and c respectively). The version with the linear slow variable tended to average way fast-timescale information, giving a similar response for both regular and burst stimulation patterns (Figure 3b). In contrast, the version with the nonlinear slow variable was able to separate the two stimulation patterns by amplifying the magnitude differences between the fast variable signals (Figure 3c). This cartoon model demonstrates that the exact tuning of nonlinearities in the fast and slow variables matter, implying further quantitative experimental and computational modelling work is needed to characterise these processes in *in vivo*-like conditions.

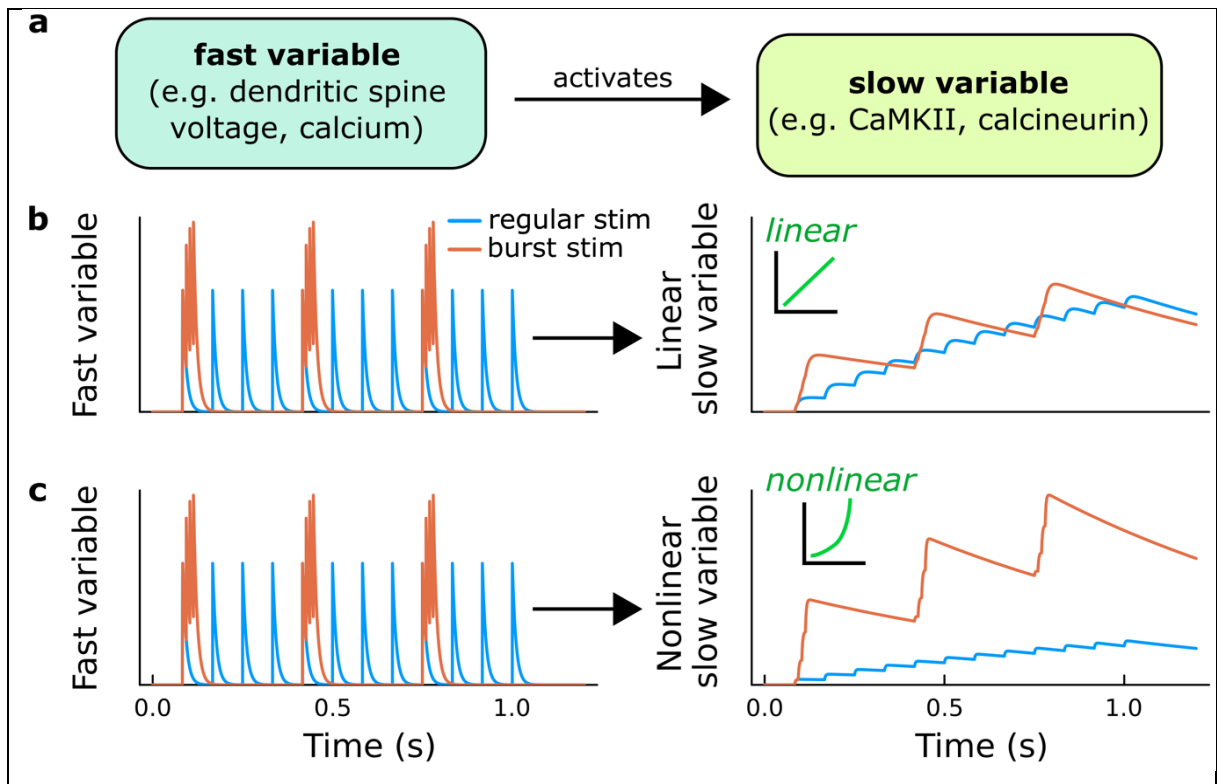

**Figure 3. Simple computational model of biochemical signalling at a synapse showing importance of slow-variable nonlinearities in plasticity induction.**

**a)** Hypothetical two-step synaptic plasticity cascade where a fast variable activates a slower variable.

**b, c)** Twelve stimulation pulses are given to the synapse model in one second, either at regular intervals (blue), or in three bursts of four (red). The fast variable (left plots) linearly sums these inputs and decays rapidly toward baseline. The slow variable's response (right plots) accumulates the signal from the fast variable, but does not distinguish the regular vs burst stimulation if linear (b), whereas if nonlinear it differentially responds more to the burst stimulus (c).

For decades, synaptic plasticity research was hampered by the crudeness of available recording and stimulation methods, but this is rapidly changing with the development of new experimental and computational tools. The synaptic plasticologist's dream experiment is now within grasp: to manipulate and monitor molecular and electrophysiological properties of individual synapses over long time periods, during behavioural learning *in vivo*. Glutamate imaging can track neurotransmitter vesicle release [73*,74], while glutamate uncaging enables focal stimulation of single synapses [75]. Correlating pre- and postsynaptic *in vivo* neural activity with synaptic properties over time can give insight into plasticity rules [27*,29**,76–78]. And probably most crucially for understanding slow-timescale variables in plasticity, advances in FRET and FLIM-based imaging of protein activity allows unprecedented measurement of the dynamics of biochemical signalling at synapses and dendrites [79]. On the computational front, computer-vision methods allow for automatic analysis of large imaging datasets [80,81]. Although running computer models with dynamics across multiple timescales is a notoriously difficult issue, advanced numerical simulation packages for differential equations and stochastic systems are now available in the Julia programming language [82]. The parameters of these biophysical models can be fit directly to data using modern automatic differentiation and gradient-descent optimisation methods [83,84*]. Machine-learning based approaches can also be used to infer abstract plasticity rules directly from neural activity during learning [85*–87]. These experimental and computational techniques can now be used hand-in-hand to discover the principles of brain learning across timescales.

How should future computational modelling work on this problem best proceed? One consideration is the degree of model realism vs abstraction. Of course, the goal of any modelling project should not be to recapitulate the full details of real brains, but to answer a specific question. As a result the appropriate degree of 'realism' for any model depends on the question at hand. For example if the question is: 'what are the different roles of L-type vs R-type calcium channels in synaptic plasticity induction?' then the model will need to formulated in appropriately biophysically-grounded dynamic equations. In contrast, a more simple abstract model may be useful for asking a question like: 'which plasticity rule better captures our physiology data, spike-timing-dependent plasticity or behavioural-timescale plasticity'? Another key aspect of the model-realism decision is practical. Detailed synapse models are slow to simulate on a computer and carry many parameters to set. Both these properties make them unwieldy for use in large simulations of many neurons interacting a circuit, especially when studying processes over multiple timescales. The ideal scenario would be to start with a biophysically-grounded model which is then reduced to a simpler more practical form [88–91] – this would give us faith that the simple model has a solid biological basis. The process of model reduction can be done via either generic rigorous mathematical techniques with controlled approximation errors, such as separation-of-timescales [92,93], or via tasteful domain knowledge; for example if we knew of a behavioural pharmacology experiment that found that kainite receptors do not contribute to the memory process we were attempting to study, then we might drop them from our synaptic plasticity model.

A second consideration is the type of question being asked, which we may generally class as mechanistic vs normative. Mechanistic questions ask *how* things work, normative questions ask *why* they might work that way. Note that mechanistic does not necessarily imply realistic: what counts as a mechanism may depend on the question, researcher, or field. Open mechanistic questions about slow-timescale factors in synaptic plasticity include: which molecular mechanisms at synapses integrate learning signals over slow timescales? How do the dynamics of short-term synaptic plasticity affect long-term plasticity induction? Which timescales of neural activity most contribute to synaptic plasticity *in vivo*? In contrast, normative questions aim to discover if we can find evidence for a particular theory of brain function, often appealing to some optimality criteria. Example normative questions on slow-timescale plasticity might include: how should brains optimally combine learned information across fast and slow timescales? How should the timescales of synaptic plasticity vary depending on sensory modality or behavioural task? What information do non-Hebbian plasticity rules such as BTSP 'learn' from neural activity?

Given these considerations, one bottom-up strategy for computational modelling of slow-timescale plasticity might be as follows: 1) design a biophysically-grounded model of synaptic plasticity that is constrained to experimental data for a given synapse type; 2) reduce the model to a simpler form to aid computer simulation, parameter fitting, understanding, and mathematical analysis; 3) simulate the simple plasticity rule in neural circuit models; 4) look for 'signatures' of the model's predictions in *in vivo* physiological or behavioural data; 5) analyse the model's properties through the lens of a normative theory, to discover what purpose these slow-timescale factors may be serving for brain learning.

**Acknowledgements**

This work was funded by grants from the Medical Research Council (MR/S026630/1), Leverhulme Trust (RPG-2019-229), and Biotechnology and Biological Sciences Research Council (BB/W001845/1).

This study used FRET imaging to quantify CaMKII phosphorylation dynamics in single dendritic spines. The authors then developed a quantitative computational model that could account for the data, which will be useful for future plasticity modelling studies.